\documentclass{article}

\usepackage{arxiv}

\usepackage[utf8]{inputenc} 
\usepackage[T1]{fontenc}    
\usepackage{hyperref}       
\usepackage{url}            
\usepackage{booktabs}       
\usepackage{amsfonts}       
\usepackage{nicefrac}       
\usepackage{microtype}      
\usepackage{lipsum}     
\usepackage{graphicx}
\usepackage{natbib}
\usepackage{doi}
\usepackage{caption} 
\usepackage{longtable}
\usepackage{pdfpages} 
\usepackage{amsmath}

\hypersetup{hidelinks}

\title{Quality Perceptions and Intended Engagement in Response to AI-Generated and AI-Assisted News}

\author{
    Fabrizio Gilardi
    \thanks{Corresponding author. Email: \href{mailto:gilardi@ipz.uzh.ch}{gilardi@ipz.uzh.ch}.}\\
    University of Zurich\\
    \And
    Sabrina Di Lorenzo \\
    University of Zurich\\
    \And
    Juri Ezzaini \\
    University of Zurich\\
    \And
    Beryl Santa \\
    University of Zurich\\
    \And
    Benjamin Streiff \\
    University of Zurich\\
    \And
    Eric Zurfluh \\
    University of Zurich\\
    \And
    Emma Hoes\\
    University of Zurich\\
}

\date{forthcoming, \emph{Scientific Reports}\\ \vspace{3pt} \today}

\renewcommand{\headeright}{}
\renewcommand{\undertitle}{}
\renewcommand{\shorttitle}{Quality Perceptions and Intended Engagement in Response to AI-Generated and AI-Assisted News}

\hypersetup{
pdftitle={A template for the arxiv style},
pdfauthor={Fabrizio Gilardi},
pdfkeywords={Artificial Intelligence, News},
}

\begin{document}
\maketitle

\begin{abstract} 

The increasing use of artificial intelligence (AI) in news production raises important questions about how audiences perceive and respond to AI-generated journalism. This preregistered survey experiment (N = 599, German-speaking Switzerland) examines (i) perceptions of article quality (measured as credibility, readability, and expertise) across news excerpts that were human-written, AI-assisted, or fully AI-generated, and (ii) self-reported intentions to engage following disclosure of AI involvement. Participants rated two short news excerpts before learning how they had been produced. Articles across all conditions were evaluated similarly in perceived quality. After disclosure, participants in the AI-assisted and AI-generated conditions reported a higher willingness to continue reading their assigned articles compared to the control group, but future willingness to read AI-generated news did not differ across conditions. Overall, the findings suggest that readers assess AI-generated and human-written news comparably in quality, while disclosure of AI use can momentarily increase curiosity or interest without yet changing longer-term reading intentions.

\end{abstract}

\newpage
\section{Introduction}

The news media landscape is evolving significantly with the advancement of artificial intelligence (AI). AI applications, especially in natural language generation, have become integral to the production of news \citep{Cools:2024a,Simon:2024c,Simon:2024b}. The integration of AI in the news presents both opportunities and risks for democracy \citep{Arguedas:2024a}. On the one hand, AI has the potential to make news more accessible and tailored to diverse audiences, supporting informed public debate. On the other hand, challenges arise around transparency, bias, and trust, as AI models often reflect the dominant worldviews of their creators, potentially narrowing the range of perspectives in public discourse. Previous research has found that audiences tend to distrust AI-generated content, particularly in politically sensitive contexts, and often prefer human-authored news, even when AI-generated content is evaluated as equally credible \citep{Waddell:2018a,Toff:2024a,Jia:2024b}.

A central issue concerns how audiences perceive and respond to AI-produced news. Understanding these reactions is crucial because news organizations are increasingly experimenting with AI tools while public skepticism remains high. In Switzerland, for instance, only 29\% of respondents would read fully AI-generated news, 55\% would read news if AI assisted in its creation, and 84\% would prefer news created without AI involvement \citep{vogler}. Similar reservations have been observed in other countries \citep{Fletcher:2024a}. These findings suggest lingering doubts about AI's role in journalism, but they do not reveal why such skepticism persists. Prior work often assumes that aversion stems from perceptions of lower quality \citep{Waddell:2018a}, yet few studies have directly compared perceived quality of AI-generated and human-written news when readers are unaware of authorship. Moreover, most existing evidence predates large language models (LLMs), whose linguistic capabilities differ substantially from earlier forms of automated journalism.

This study addresses these gaps by examining both perceptions of article quality and intentions to engage with AI-generated and AI-assisted news. Using a preregistered survey experiment with Swiss participants, we compare evaluations of short news excerpts written by journalists, rewritten with AI assistance, or generated entirely by AI. Participants rated each article's credibility, readability, and expertise before learning how it was produced, allowing us to assess whether perceptions differ when authorship is unknown. We then measure self-reported willingness to continue reading after disclosure of AI involvement, as well as general willingness to read AI-generated news in the future. This design enables us to separate initial quality assessments (focused on message characteristics) from subsequent reactions influenced by source information.

Our study contributes to two ongoing debates. First, it provides up-to-date evidence on whether audiences perceive differences in the quality of human-written, AI-assisted, and AI-generated news. Second, it examines how disclosure of AI involvement affects short-term engagement intentions, recognizing that any observed increases in willingness to continue reading may reflect curiosity or skepticism rather than genuine acceptance. The findings that follow show that readers evaluate AI-generated and human-written news similarly in perceived quality, and that disclosure of AI use can briefly increase interest in the presented articles without altering broader reading intentions. Rather than implying that AI disclosure ``builds trust'' or ``enhances engagement,'' these results highlight how curiosity and heuristics about AI shape immediate reactions. Because our design measures self-reported intentions rather than actual or long-term behavior, we do not infer sustained acceptance of AI-generated news. Instead, we discuss how these short-term responses contribute to understanding the evolving relationship between AI transparency, reader perceptions, and the credibility of journalistic content.

\section{Evolving Perceptions of AI in News Production}

AI is increasingly integrated into news production workflows, handling tasks such as data analysis, trend identification, content aggregation, and personalization \citep{Cools:2024b}. Early forms of automated journalism were largely limited to structured, data-based formats such as sports or financial reporting, where template-based systems could produce standardized text. These early systems demonstrated AI's potential to handle repetitive and data-heavy tasks, freeing journalists to focus on work that demands human judgment, creativity, and contextual reasoning, such as investigative reporting and nuanced writing \citep{clerwall,haigra}. However, the introduction of large language models (LLMs) such as GPT-type architectures has expanded AI's linguistic capabilities far beyond rule-based automation. LLMs can now generate fluent, stylistically adaptive, and contextually coherent news-like text, raising qualitatively different questions about authorship, accountability, and trust \citep{Cools:2024b,Simon:2024b}. 

Research on public perceptions of AI-generated news presents a mixed picture. Earlier studies of automated journalism, focused on algorithmic or template-based reporting, found that readers often preferred human-written articles \citep{graefe2018,graboh}, yet more recent experiments show minimal or no differences in perceived credibility, expertise, and readability between human-written and AI-generated news \citep{wölpow,kim2020,kim2022}. A Turing test-based study on Spanish-language AI-generated journalism similarly concluded that AI-authored pieces can be perceived as equally credible as those written by professional journalists, even if human reporters are still regarded as more engaging storytellers \citep{Barrolleta:2024a}. Newer research on generative AI extends these findings to more complex, less formulaic domains: for instance, audience trust in LLM-generated political or cultural reporting appears highly contingent on disclosure cues and perceived human oversight \citep{Toff:2024a,Jia:2024b,Nanz:2025a,Gondwe:2025a}. Across these studies, readers tend to judge AI-authored content as technically competent but remain uncertain about the intentions, transparency, and accountability behind it. 

Attribution and disclosure continue to play a pivotal role in shaping perceptions of credibility. Research indicates that many readers overestimate the degree of automation involved, assuming full algorithmic authorship even when AI contributions are minor \citep{Bien-Aime:2025a,altgil}. The manner in which AI involvement is presented (whether as full authorship, partial assistance, or editorial support) strongly affects perceived message and source credibility \citep{Jia:2024b}. Disclosure about AI authorship can reduce trust and ad acceptance even when content evaluations are equivalent \citep{Nanz:2025a}, whereas transparency and clear labeling can build trust in contexts where digital literacy and local relevance are high \citep{Gondwe:2025a}. These findings suggest that disclosure effects are not uniformly positive or negative; rather, they interact with audience heuristics, prior attitudes, and the perceived legitimacy of AI use in journalism.

While much of this literature focuses on credibility perceptions, a smaller body of work has examined how audiences' \emph{willingness to engage} with AI-produced news responds to disclosure. Survey evidence from Switzerland shows that stated willingness to read AI-generated news is substantially lower than for human-written or AI-assisted news, even before any exposure to actual content \citep{vogler}. Cross-national data confirm this pattern, with majorities expressing discomfort with fully automated news across diverse media systems \citep{Fletcher:2024a}. At the same time, experimental work suggests that stated reluctance does not always translate into avoidance behavior: once exposed to AI-generated content, audiences may update their engagement intentions, at least temporarily \citep{Toff:2024a}. Recent research on willingness to pay for AI-generated news further indicates that engagement decisions are shaped not only by perceived quality but also by normative expectations about journalistic labor and human oversight \citep{Nanz:2025a}. These findings highlight the need to study engagement intentions alongside credibility perceptions, as the two do not necessarily move in parallel.

Bias perceptions remain a central concern. While some studies suggest that audiences view AI-generated reporting as more objective than human journalism \citep{wölpow,Wang:2024e}, this perception may be fragile. AI systems trained on historical data may replicate existing biases while giving the impression of neutrality. Experiments indicate that audiences evaluate AI-generated articles as less biased only when AI involvement is undisclosed, implying that perceived objectivity depends on ignorance of authorship rather than on actual algorithmic fairness \citep{Waddell:2019a,Waddell:2018a}. Newer generative-AI research similarly finds that disclosure can transform perceived neutrality into skepticism, as audiences question whose values and assumptions underlie machine-generated text \citep{Toff:2024a,Gondwe:2025a}. 

Transparency is thus a double-edged principle: while often promoted to foster accountability, labeling AI authorship may paradoxically lower trust \citep{Wang:2024d,Wang:2024e}. Readers frequently express doubt toward content labeled as AI-generated, regardless of whether the text is actually machine- or human-written, reflecting a broader ``automation heuristic'' in which AI attribution triggers defensive skepticism \citep{altgil}. Conversely, when AI involvement is not disclosed, credibility assessments tend to converge with those for human-written news \citep{Jia:2024b,Waddell:2018a}. This asymmetry underscores the persistence of audience heuristics about machine agency and authenticity even in the era of sophisticated generative systems.

The shift to LLM-driven journalism raises new challenges for responsible integration. Issues of algorithmic bias, misinformation, and ``hallucinated'' content remain salient, as does the need to safeguard journalistic autonomy and editorial accountability \citep{Cools:2024b}. Recent findings show that most news organizations still lack standardized policies for attribution and internal oversight \citep{Bien-Aime:2025a}, while others call for frameworks ensuring that AI tools assist rather than replace human editors \citep{Araujo:2023a,Cools:2024a,Dodds:2024a}. As generative AI systems are adopted globally, cross-cultural studies such as \cite{Gondwe:2025a} remind us that local media ecologies and cultural expectations about fairness and voice strongly shape audience responses to AI-generated journalism.  

Overall, existing research converges on three key insights. First, contemporary LLM-based journalism differs fundamentally from early automated approaches by operating in open-ended linguistic domains rather than fixed data templates. Second, perceptions of credibility and trust depend not only on textual quality but also on contextual signals of human oversight and transparency. Third, most prior studies have not disentangled how readers evaluate content quality before disclosure from how they respond once authorship is revealed. Our study extends this literature by experimentally isolating these stages: assessing perceived quality without knowledge of authorship and measuring subsequent engagement intentions after disclosure. In doing so, it bridges findings from earlier automated journalism to the emerging reality of generative AI, offering evidence on how contemporary audiences perceive and react to human-AI collaboration in news production.

\section{Research Questions and Hypotheses}

Building on the existing literature and the gaps identified in previous studies, our research aims to systematically examine how AI involvement in news generation affects audience perceptions. Specifically, we address the following research questions:

\begin{description}
    \item[RQ1:] Prior to disclosing AI's role, how do respondents evaluate the quality of AI-assisted and AI-generated news articles in terms of expertise, readability, and credibility, compared to human-written news articles?
    \item[RQ2:] How does disclosing AI's role in generating an article influence respondents' willingness to engage with AI-assisted or AI-generated news after they have rated the article's quality?
\end{description}

These questions allow us to extend prior research in three ways. First, we operationalize ``quality'' as a multidimensional construct, assessing expertise, readability, and credibility across three distinct levels of AI involvement, reflecting real-world journalistic practices. This operationalization follows the framework established by \citet{sundar1999} and applied in prior automated journalism research \citep{haigra,graefe2018}, which distinguishes between the technical competence of the writing (expertise), its stylistic appeal (readability), and the perceived trustworthiness of the content (credibility). Second, instead of relying on abstract survey questions about AI, which may be influenced by preconceptions, we ground participants' responses in concrete examples of AI-generated and AI-assisted news. Third, we separate the evaluation of news article quality from attitudes toward AI-generated news by only revealing AI's involvement \textit{after} respondents have rated the articles. This design ensures that initial quality assessments are not biased by prior knowledge of AI authorship.

\subsection{Quality Perceptions of AI-Generated News}

The first set of hypotheses investigates whether AI-assisted or AI-generated news is perceived as lower in quality compared to human-written articles. Prior studies have found that human-written news is generally rated higher for quality \citep{graboh}. This aligns with the authority heuristic and social presence heuristic, which suggest that content created by humans is typically seen as more credible due to its perceived human expertise and social presence \citep{sundar2008}. Given these established biases, we expect that articles with AI involvement will receive lower quality ratings, though more recent findings challenge this assumption \citep{wölpow,Barrolleta:2024a}.

\begin{description}
    \item[H1:] Respondents in the AI-assisted condition rate the article's \underline{expertise} lower than respondents in the human-written condition.
    \item[H2:] Respondents in the AI-assisted condition rate the article's \underline{readability} lower than respondents in the human-written condition.
    \item[H3:] Respondents in the AI-assisted condition rate the article's \underline{credibility} lower than respondents in the human-written condition.
    \item[H4:] Respondents in the AI-generated condition rate the article's \underline{expertise} lower than respondents in the human-written condition.
    \item[H5:] Respondents in the AI-generated condition rate the article's \underline{readability} lower than respondents in the human-written condition.
    \item[H6:] Respondents in the AI-generated condition rate the article's \underline{credibility} lower than respondents in the human-written condition.
\end{description}

\subsection{Impact of AI Disclosure on Engagement}

The second set of hypotheses examines how awareness of AI involvement affects engagement with the article. Previous research suggests that AI-generated content is often met with skepticism, particularly when it covers political topics \citep{Wang:2024d,Wang:2024e,vogler,tandoc}. We therefore hypothesize that respondents will be less inclined to continue reading once they learn about AI's role. However, recent findings that disclosure may also spark curiosity or novelty-driven engagement may challenge this premise \citep{Toff:2024a}.

\begin{description}
    \item[H7:] Respondents in the AI-assisted condition express a lower willingness to \underline{continue reading} the article than respondents in the human-written condition.
    \item[H8:] Respondents in the AI-generated condition express a higher willingness to \underline{continue reading} the article than respondents in the human-written condition.
\end{description}

\subsection{Long-Term Attitudes Toward AI-Generated News}

The final set of hypotheses assesses how engagement with AI-generated content influences respondents' general willingness to read AI-generated news in the future. Unlike the previous hypotheses, which examine reactions to specific articles, these hypotheses explore broader attitudes towards AI-generated journalism. While prior surveys indicate high levels of skepticism towards AI-generated news \citep{vogler}, experimental research suggests that direct exposure to AI-generated content may mitigate these concerns, as familiarity with AI often reduces skepticism and increases trust \citep{novozhilova}.

\begin{description}
    \item[H9:] Respondents in the AI-assisted condition express a higher willingness to \underline{read AI-generated news in the future} than respondents in the human-written condition.
    \item[H10:] Respondents in the AI-generated condition express a higher willingness to \underline{read AI-generated news in the future} than respondents in the human-written condition.
\end{description}

By disentangling quality assessments from short-term engagement dynamics and long-term attitudes, our study contributes to a more detailed understanding of how AI-generated journalism is perceived.


\subsection{Comparison Between AI-Assisted and AI-Generated Conditions}

In addition to the preregistered hypotheses comparing AI-assisted and AI-generated articles to human-written ones, we also preregistered exploratory questions examining differences between the two AI conditions. These questions assessed potential contrasts in perceived expertise, readability, and credibility, as well as in willingness to continue reading and willingness to read AI-generated news in the future. The corresponding results are reported in Appendix \ref{appendix:questions} and show no significant differences between AI-assisted and AI-generated articles across any of these measures.

\section{Experimental Design}

To explore the public's perceptions of AI-generated and AI-assisted news articles compared to traditional human-written articles, we conducted a pre-registered online between-subjects survey experiment.\footnote{The pre-analysis plan is available at: \url{https://osf.io/vw8a3}.} We recruited 599 participants from the age of 18 from the German-speaking part of Switzerland through the survey company Bilendi, using quotas based on age and gender to ensure balance across these dimensions. The study was conducted in May 2024. Sample size was determined by a power analysis, assuming an effect size of Cohen's $d$ = 0.3 for the primary outcomes (perceived quality dimensions) and $d$ = 0.15 for the secondary outcomes (engagement intentions). On the 1--5 outcome scales used in this study, these correspond to expected regression coefficients of approximately 0.17--0.20 and 0.08--0.10, respectively (see below for details).

\begin{figure*}
\centering
\resizebox{\textwidth}{!}{%
\includegraphics{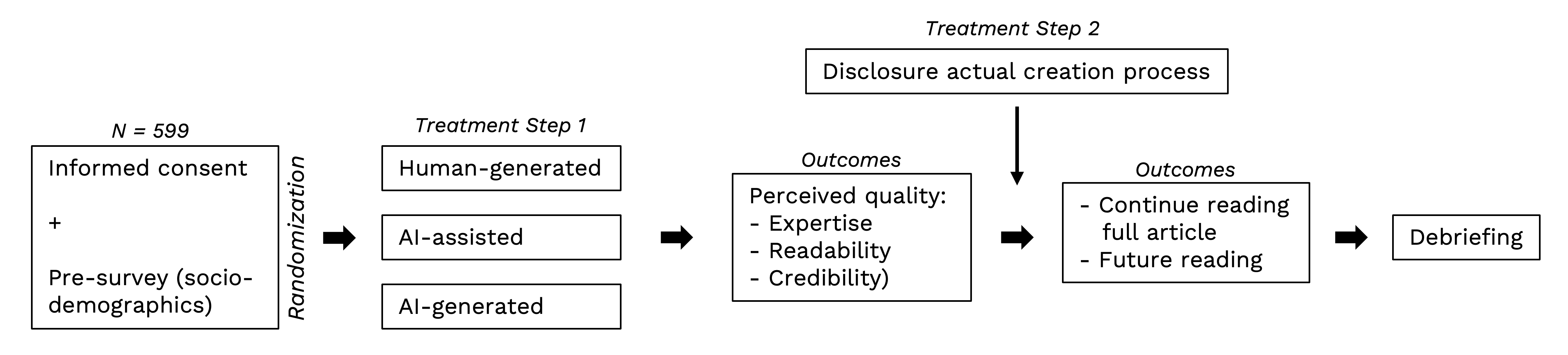}
}
\caption{Survey experiment flow}
\label{flow}
\end{figure*}

Figure \ref{flow} depicts the survey experiment flow: After informed consent, participants first completed a pre-survey to determine their socio-demographic characteristics, including age, gender, education level, and political orientation. Participants were then randomly assigned to one of three conditions: a control group that read excerpts from human-written articles (``human-written'' condition), a first treatment group that read excerpts that were rewritten with the help of ChatGPT (``AI-assisted'' condition), and another treatment group that read excerpts that were entirely generated by ChatGPT (``AI-generated'' condition).\footnote{Throughout this paper, we use ``human-written condition'' and ``control group'' interchangeably to refer to the same experimental condition.}

Socio-demographic variables were measured as follows. \emph{Age} was recorded as a continuous variable in years. \emph{Gender} was measured with five response options (female, male, non-binary, other, prefer not to answer); for the analysis, it was coded as a binary variable (0 = male, 1 = female). \emph{Education} was measured on a four-level ordinal scale: 0 = compulsory schooling (\textit{Obligatorische Schule}), 1 = vocational apprenticeship (\textit{Berufslehre}), 2 = upper secondary or higher vocational education (\textit{Matura, Höhere Berufsbildung}), 3 = university or university of applied sciences (\textit{Universität, Fachhochschule}). \emph{Political orientation} was measured on a five-point ordinal scale: 0 = left (\textit{Links}), 1 = center-left (\textit{Eher links}), 2 = center (\textit{Mitte}), 3 = center-right (\textit{Eher rechts}), 4 = right (\textit{Rechts}), with an additional response option ``don't know / prefer not to say'' (\textit{Keine Antwort / Weiss nicht}) coded as missing. In the regression models, higher values on each variable correspond to older age, female gender, higher education, and more right-leaning political orientation, respectively. A randomization balance check confirmed no significant differences across conditions for any of these variables (all $F < 1.55$, all $p > .21$; see Appendix~\ref{appendix:balance}).

To ensure consistency and create a set of texts that is comparable across the three groups, all articles are derived from actual articles on Swiss politics published on the news website of Switzerland's public broadcaster (SRF), which ensures homogeneity in style, quality, as well as topic to a certain extent (that is, only Swiss politics, with variation in sub-topics). These articles are shown to the control group. We used the following procedure to generate articles with AI involvement. For AI-assisted articles, we copy-pasted the original article into ChatGPT and asked it to rewrite the article without losing any information. This resulted in articles for the AI-assisted group that are similar to the originals, but are rewritten by the AI system. For AI-generated articles, we only provided ChatGPT with the title and lead of the original article. We then asked ChatGPT to generate a short article in the same style as the source of the human-written articles. English translations of example stimuli are shown in Appendix \ref{appendix:stimuli}; the original German versions are available in the replication materials.

Each participant read two excerpts, randomly drawn from a pool of ten articles for each category (human-written, AI-assisted, AI-generated), making a total pool of 30 articles across all experimental conditions. Each excerpt had a hard cutoff at 150 words to maintain consistency, avoid bias due to article length, and simulate a paywall for the purposes of our third outcome (willingness to ``continue reading'' the article). The random selection of two articles from a pool of ten texts for each group minimizes the risk that the outcomes depend on specific topics. Moreover, for each article, we asked respondents to answer a simple question on the article's content. This allows us to check whether respondents have read the text in sufficient detail.

The first outcome measures how participants evaluate the quality of the articles. Following \citet{sundar1999} and as preregistered, we asked respondents to rate three dimensions of quality: \textit{expertise}, \textit{readability}, and \textit{credibility}. Each dimension is based on specific items, expressed as adjectives \citep{haigra}: ``clear,'' ``coherent,'' ``comprehensive,'' ``concise,'' and ``well-written'' for expertise; ``boring,'' ``enjoyable,'' ``interesting,'' ``lively,'' and ``pleasing'' for readability; and ``biased,'' ``fair,'' and ``objective'' for credibility. Participants rated the articles on each of these items using a 1–5 Likert scale, and we then aggregated the scores for each dimension. Conceptually, while our preregistration labeled ``well-written'' as measuring \textit{expertise}, we acknowledge that it also captures \textit{readability}, reflecting stylistic clarity rather than authorial competence. Moreover, we acknowledge that our operationalization does not encompass broader facets of perceived news quality \citep[e.g.,][]{Urban:2014a,Appelman:2016a}, which future research could incorporate to strengthen construct validity.

Following our preregistered procedure, we verified that the three quality dimensions (expertise, readability, and credibility) formed a reliable scale (Cronbach's $\alpha \geq$ 0.70). While our main analyses rely on the separate dimensions, we reran all models using the aggregate quality index; the results are substantively identical (see Appendix \ref{appendix:index}).

In a second step, following the initial quality rating, participants were informed about the actual creation process of the articles (the exact wording of the disclosures is shown in Appendix \ref{appendix:disclosures}). After learning how the texts they read were created, respondents were asked if they would be willing to continue reading the full articles after reviewing the excerpts again. In addition, all respondents were asked to report if they would be willing to read AI-generated news articles in the future, on a 1-5 Likert-scale. A key feature of this design is that participants learned about AI involvement only after rating article quality, avoiding bias in initial assessments that labeling can introduce \citep{Toff:2024a,Jia:2024b,Waddell:2018a}. This sequencing prioritizes internal validity by isolating the effect of disclosure on engagement intentions, though it departs from real-world settings where disclosure typically precedes exposure.

As preregistered, we estimated linear regression models for each dependent variable with treatment condition as a three-level categorical factor (human-written, AI-assisted, AI-generated) and demographic covariates (age, gender, education, political orientation). This specification is algebraically equivalent to a one-way ANCOVA: the omnibus $F$-test for the treatment factor tests whether condition means differ, while planned pairwise contrasts directly test the preregistered hypotheses (H1--H10). The regression framework accommodates the preregistered demographic controls and provides coefficient estimates that are directly interpretable as adjusted mean differences. To further verify that specific article topics did not drive the results, we re-estimated all models with article-pair fixed effects; the findings are unchanged (see Appendix~\ref{appendix:robustness}).

The preregistered power analysis specified expected effect sizes as Cohen's $d$. Since $d = b / \text{SD}_{\text{pooled}}$, the preregistered $d$ = 0.30 for primary outcomes corresponds to expected regression coefficients of approximately 0.17--0.20 on the 1--5 quality scales (SD $\approx$ 0.55--0.66), and $d$ = 0.15 corresponds to approximately 0.08--0.10. For secondary outcomes (engagement intentions, SD $\approx$ 1.16--1.20), $d$ = 0.30 corresponds to coefficients of approximately 0.35. To facilitate interpretation of effect magnitudes, we report Cohen's $d$ for all pairwise comparisons and $\eta^2$ and $\omega^2$ for the omnibus tests, alongside significance tests and the preregistered Benjamini--Hochberg adjustment for multiple comparisons.

All pairwise $p$-values are reported alongside Benjamini--Hochberg adjusted $p$-values. For ease of presentation, the main tables show both unadjusted and adjusted significance levels; full details are provided in Appendix~\ref{appendix:BH}.

\section{Results}

Figure \ref{quality} presents the mean ratings for the first outcome, assessing perceived article quality across credibility, expertise, and readability. Across all three groups (human-written, AI-assisted, and AI-generated) the mean scores were highly similar, with overlapping confidence intervals indicating no significant differences. In other words, articles written by journalists, rewritten with AI assistance, or generated entirely by AI were perceived as equally credible, readable, and expert.

\begin{figure*}
\centering
\resizebox{\textwidth}{!}{%
\includegraphics{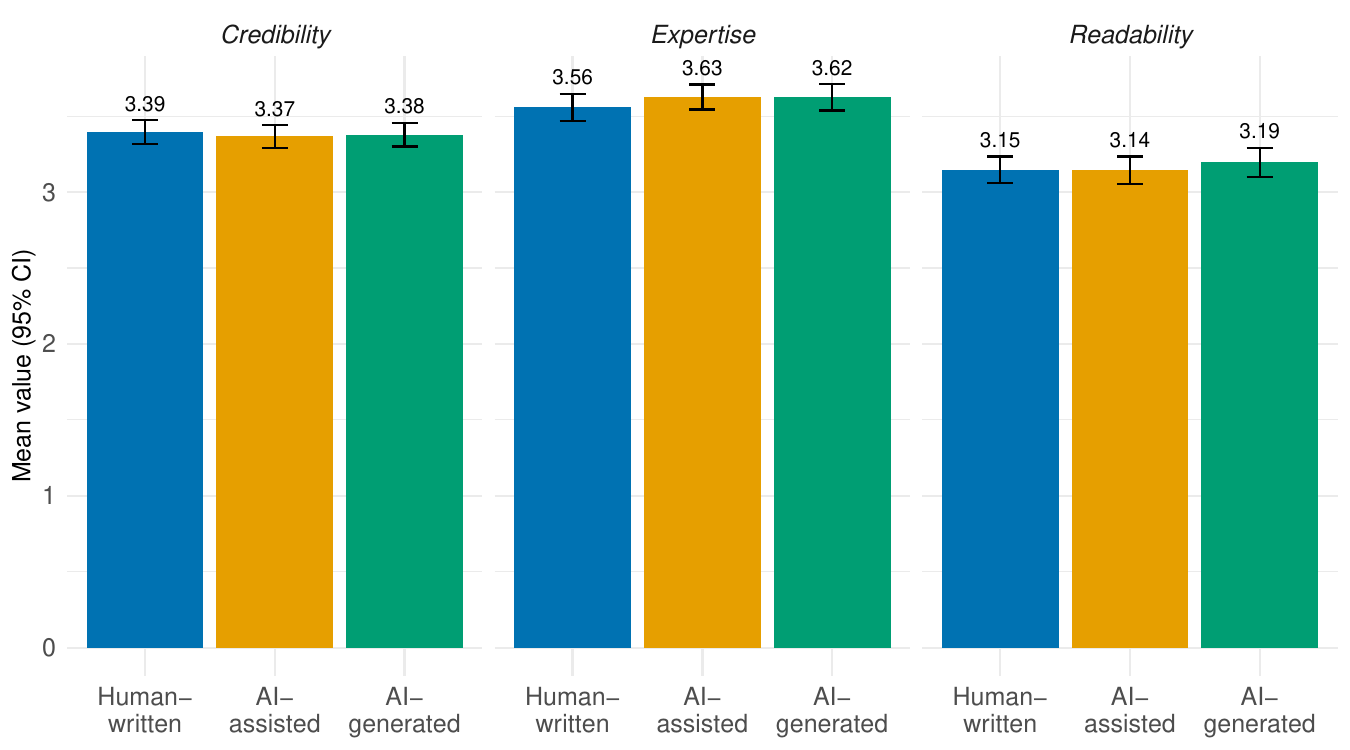}
}
\caption{Perceived article quality (credibility, expertise, readability) across human-written, AI-assisted, and AI-generated news. Mean ratings (1–5) were collected before disclosure of AI involvement.}
\label{quality}
\end{figure*}

\begin{figure*}
\centering
\resizebox{\textwidth}{!}{%
\includegraphics{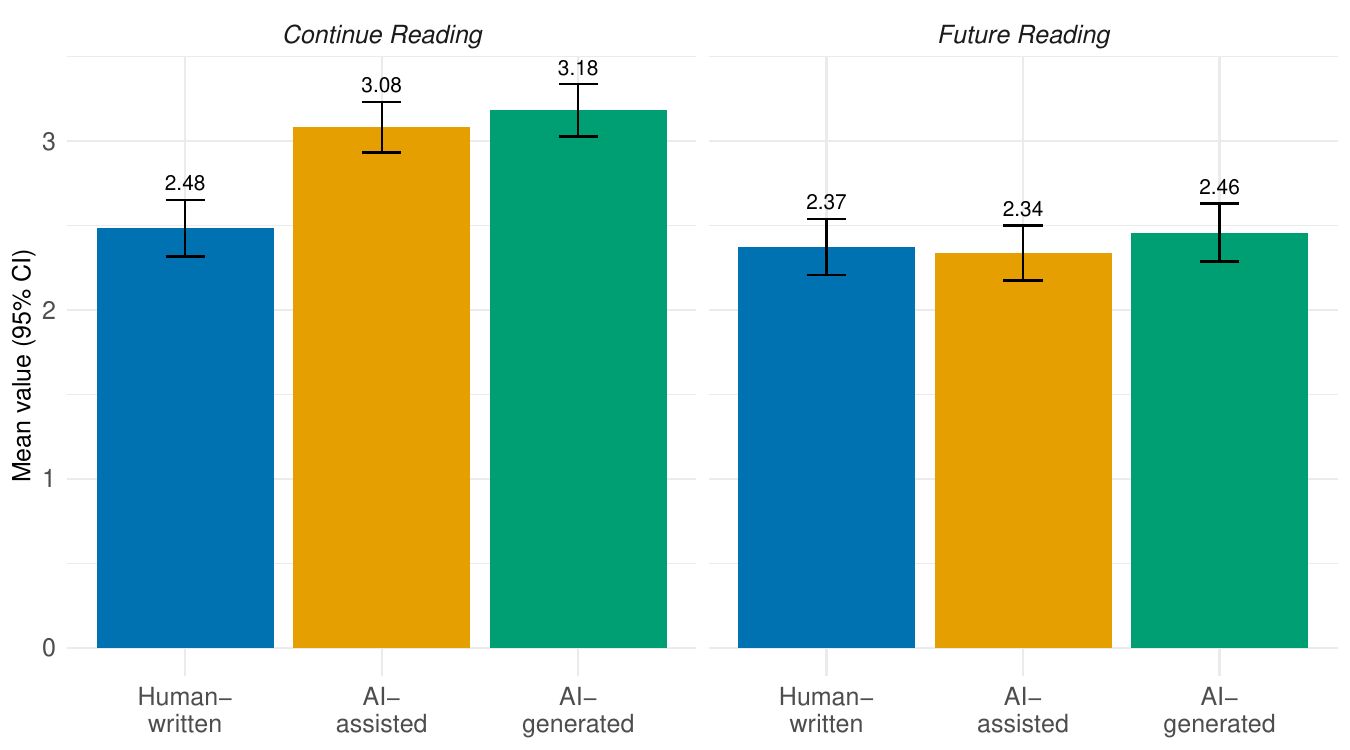}
}
\caption{Engagement intentions (1-5) after disclosure of AI involvement. Participants in AI-assisted and AI-generated conditions showed higher willingness to continue reading but similar future willingness to read AI-generated news, suggesting short-term curiosity rather than lasting acceptance.}
\label{read}
\end{figure*}

Table~\ref{tab:singlefactor} presents the primary analysis: linear regression models with treatment as a three-level factor and demographic covariates, estimated separately for each dependent variable. For all three quality dimensions (expertise, readability, and credibility), neither the AI-assisted nor the AI-generated condition differs significantly from the human-written condition. The omnibus $F$-tests for the treatment factor are non-significant across all quality outcomes (all $p > .45$; $\eta^2 \leq .003$, $\omega^2 \approx 0$), and the corresponding pairwise Cohen's $d$ values are negligible (ranging from $-$0.05 to 0.11). These results remain unchanged when estimating the models only for the subset of respondents who passed the manipulation check.

\begin{table} \centering 
  \caption{Regression results with treatment as a three-level factor. Coefficients show differences relative to the human-written condition.} 
  \label{tab:singlefactor}  
\begin{tabular}{@{\extracolsep{3pt}}lccccc} 
\hline \\[-1.8ex] 
 & Expertise & Readability & Credibility & Continue  & Future \\ 
 &  &  &  &  Reading &  Reading \\ 
\hline \\[-1.8ex] 
 AI-assisted & 0.060 & $-$0.00002 & $-$0.014 & 0.566$^{***}$ & 0.010 \\ 
  & (0.066) & (0.070) & (0.060) & (0.121) & (0.124) \\ 
  & [$d$ = 0.11] & [$d$ = $-$0.01] & [$d$ = $-$0.05] & [$d$ = 0.53] & [$d$ = $-$0.03] \\[4pt]
 AI-generated & 0.073 & 0.048 & $-$0.009 & 0.697$^{***}$ & 0.164 \\ 
  & (0.067) & (0.072) & (0.061) & (0.123) & (0.127) \\ 
  & [$d$ = 0.11] & [$d$ = 0.07] & [$d$ = $-$0.03] & [$d$ = 0.61] & [$d$ = 0.07] \\[4pt]
 Age & 0.002 & 0.005$^{**}$ & 0.001 & 0.003 & $-$0.016$^{***}$ \\ 
  & (0.002) & (0.002) & (0.002) & (0.003) & (0.003) \\ 
 Female & 0.016 & $-$0.024 & $-$0.061 & $-$0.042 & $-$0.284$^{**}$ \\ 
  & (0.055) & (0.059) & (0.051) & (0.101) & (0.105) \\ 
 Education & 0.018 & $-$0.024 & 0.014 & 0.045 & 0.120$^{*}$ \\ 
  & (0.030) & (0.032) & (0.028) & (0.055) & (0.057) \\ 
 Political orientation & $-$0.040 & $-$0.038 & $-$0.048 & $-$0.049 & $-$0.108$^{*}$ \\ 
  & (0.029) & (0.031) & (0.026) & (0.052) & (0.054) \\ 
 Constant & 3.552$^{***}$ & 3.141$^{***}$ & 3.478$^{***}$ & 2.467$^{***}$ & 2.947$^{***}$ \\ 
  & (0.117) & (0.125) & (0.107) & (0.214) & (0.221) \\ 
\hline \\[-1.8ex] 
Observations & 530 & 530 & 530 & 526 & 529 \\ 
R$^{2}$ & 0.010 & 0.019 & 0.010 & 0.069 & 0.075 \\ 
Adjusted R$^{2}$ & $-$0.001 & 0.008 & $-$0.001 & 0.059 & 0.064 \\ 
$\eta^2$ (treatment) & 0.003 & 0.001 & 0.000 & 0.070 & 0.002 \\ 
$\omega^2$ (treatment) & 0.000 & 0.000 & 0.000 & 0.066 & 0.000 \\ 
\hline 
\hline \\[-1.8ex] 
\textit{Note:}  & \multicolumn{5}{l}{$^{*}p<.05$; $^{**}p<.01$; $^{***}p<.001$. Standard errors in parentheses.} \\ 
 & \multicolumn{5}{l}{Cohen's $d$ in brackets. All treatment $p$-values retain the same significance} \\ 
 & \multicolumn{5}{l}{status after Benjamini--Hochberg adjustment (see Appendix~\ref{appendix:BH}).} \\ 
\end{tabular} 
\end{table}

Consequently, we reject hypotheses H1--H6, which expected lower perceptions of expertise, readability, and credibility in the AI-assisted and AI-generated conditions compared to the human-written condition. In other words, news articles generated either with the assistance of AI or entirely by AI are perceived to match the quality of traditional articles written by journalists. This finding supports the view that AI involvement does not reduce message credibility (that is, perceptions of the content itself) even though broader aspects such as source credibility or media trust were not directly measured here.

Figure \ref{read} shows the mean ratings for the second set of outcomes, measuring participants' willingness to continue reading the presented article and to read AI-generated news in the future. Participants who learned about AI involvement—whether assisted or fully generated—were significantly more willing to continue reading than those in the human-written condition, while no difference emerged between the two AI conditions. In contrast, willingness to read AI-generated news in the future did not differ significantly across groups.

The regression results in Table~\ref{tab:singlefactor} confirm a higher willingness to continue reading in the AI-assisted ($b$ = 0.566, $p < .001$, $d$ = 0.53) and AI-generated ($b$ = 0.697, $p < .001$, $d$ = 0.61) conditions. These are moderate-to-large effects, substantially exceeding the preregistered threshold of $d$ = 0.30. The omnibus treatment effect for continue reading is significant ($\eta^2$ = 0.070, $\omega^2$ = 0.066). No significant differences emerge for willingness to read AI-generated news in the future ($d$ = $-$0.03 for AI-assisted; $d$ = 0.07 for AI-generated). We therefore reject hypotheses H9 and H10, which predicted higher willingness to read AI-generated news in the future.

The increase in willingness to continue reading should be interpreted with caution. Because participants rated this item after disclosure of AI involvement, the effect may reflect curiosity or skepticism toward AI-written news rather than genuine openness or approval. Participants may also have indicated a higher willingness simply because they had already started reading the article or wanted to see whether they could detect traces of AI authorship. Thus, the ``willingness to continue reading'' measure likely captures a form of curiosity-driven engagement rather than endorsement or long-term acceptance.

To determine whether the absence of significant effects truly indicates that readers perceived the articles as equally good, we conducted equivalence tests \citep[TOST;][]{Lakens:2018a} for all non-significant effects with a smallest effect size of interest (SESOI) of $\pm$0.2 on the 1-5 scales. As reported in Appendix \ref{appendix:TOST}, the results confirm that perceived expertise, readability, credibility, and overall quality are statistically equivalent across the three conditions, meaning that readers evaluated AI-generated and AI-assisted articles as being of the same quality as human-written ones. For future willingness to read AI-generated news, equivalence could not be established, meaning that small effects cannot be ruled out. This suggests that while large differences are unlikely, there may still be minor differences between conditions, leaving open the possibility of a modest positive effect on future engagement.

In summary, the data show no significant differences in perceived expertise, readability, and credibility between articles created with the help of AI, those fully generated by AI, and those written by humans. However, participants in the AI-assisted and AI-generated conditions exhibit a higher short-term willingness to continue reading the articles after disclosure. These findings suggest that AI involvement in news production does not negatively affect perceived content quality when readers are unaware of authorship, and that disclosure may trigger temporary curiosity or engagement. Yet, because our design captures self-reported intentions rather than actual behavior, and only at a single point in time, it cannot speak to longer-term effects on reading habits or trust. Overall, our study contributes to understanding source effects on message credibility rather than broader constructs such as media trust or enduring acceptance of AI-generated journalism.

\section{Discussion}

Beyond its empirical findings, this study contributes to theoretical discussions on source effects on message credibility. Our results show a partial decoupling of perceived content quality from audience acceptance: readers acknowledge AI-generated articles as comparably well written yet remain hesitant about their origin. This aligns with research on algorithmic aversion, which finds that people may resist algorithmic authorship despite recognizing its competence \citep{Dietvorst:2015a}. The findings also refine existing accounts of source credibility theory, indicating that while AI-generated texts can demonstrate high expertise, perceptions of trustworthiness and authenticity remain lower in the absence of human editorial accountability \citep{altgil}.

The observed increase in willingness to continue reading after disclosure warrants further consideration. This pattern is consistent with information-seeking responses to novel or unexpected stimuli: when participants learned that the articles they had just evaluated positively were in fact produced with AI involvement, this incongruity may have prompted a desire to re-examine the text. This interpretation aligns with \citet{altgil}, who found that AI-generated content labels trigger heightened scrutiny regardless of actual authorship, suggesting that disclosure activates an ``automation heuristic'' that increases attentional engagement. Importantly, this curiosity-driven response did not generalize to future reading intentions, which remained low across all conditions. The divergence between immediate and future engagement is consistent with the distinction between situational interest (triggered by novelty or surprise) and sustained individual interest \citep[e.g.,][]{novozhilova}, and suggests that a single exposure to AI-generated news is insufficient to shift broader attitudes toward AI in journalism. We note, however, that we did not measure participants' prior experience with LLMs or their baseline attitudes toward AI, which may moderate the novelty response. Future research could assess these factors to distinguish more precisely between novelty, skepticism, and emerging acceptance as drivers of post-disclosure engagement.

These results also contribute to ongoing debates about transparency in journalism. While transparency about AI use is often seen as a tool to build trust, our findings suggest that disclosure may instead elicit short-term curiosity without increasing long-term acceptance. Future research should examine how different disclosure framings, such as highlighting human oversight or explaining AI's limited editorial role, affect perceptions of credibility and trust.

Further studies could track changes in public attitudes over time through longitudinal designs, clarifying whether initial curiosity evolves into sustained acceptance or skepticism. Comparative research across linguistic and cultural contexts would help assess the generalizability of these findings, and experimental work could test interventions aimed at fostering informed and reflective engagement with AI-generated journalism. Finally, our results highlight the importance of integrating AI into news production in ways that maintain human accountability and transparency, balancing efficiency gains with the audience's enduring expectations of authenticity in journalism.

\section{Limitations}

While the study offers empirical insight into how readers perceive and respond to AI-generated and AI-assisted news, several limitations should be noted. First, regarding construct validity, our measures of article quality (credibility, readability, and expertise) capture important but not exhaustive dimensions of news evaluation, and they rely on self-reported perceptions rather than more recent, multidimensional quality scales. Second, the experiment's external validity is limited. Participants evaluated short excerpts in an artificial survey context, and disclosure of AI involvement occurred after initial quality assessments rather than before exposure, as would typically happen in real-world news environments. We did not re-measure perceived quality after disclosure, which limits our ability to assess how learning about AI authorship might retroactively alter quality judgments. Third, the study's scope is intentionally narrow: it focuses on short-term, self-reported reactions to AI authorship using a single national sample. Fourth, the experiment measures intended engagement rather than actual behavior. Readers' stated willingness to continue or return to AI-generated news may differ from their real-world consumption choices. Fifth, we did not measure participants' prior experience with large language models or their baseline news consumption habits. These factors may meaningfully shape how audiences evaluate and respond to AI-generated news, and their omission limits the precision of our interpretation of the engagement effects, particularly the novelty account. Finally, while we verified comprehension of article content through attention checks administered during the reading phase, we did not include a comprehension check after the disclosure step to verify that participants attended to and registered the stated source of the articles. Future research should address these limitations by incorporating behavioral measures, repeated exposure, prior AI experience assessments, post-disclosure comprehension checks, and cross-national designs to examine how AI disclosure affects engagement and trust in more ecologically valid settings.

\section{Conclusion}

This preregistered survey experiment examined how Swiss participants perceive the quality of human-written, AI-assisted, and AI-generated news excerpts, and how disclosure of AI involvement affects their willingness to engage with such content. We found no significant differences in perceived expertise, readability, or credibility across conditions, with equivalence tests confirming that readers evaluated the articles as comparable in quality. After disclosure, participants in both AI conditions reported higher willingness to continue reading the presented articles (a moderate-to-large effect, $d$ = 0.53--0.61) but this did not extend to future reading intentions. These findings indicate that contemporary AI-generated news can match human-written news in perceived message quality, while disclosure of AI involvement triggers short-term, curiosity-driven engagement rather than lasting changes in attitudes toward AI-generated journalism.

\section*{Ethics Approval}

This study was approved by the Ethics Committee of the Faculty of Arts and Social Sciences, University of Zurich (approval number 23.10.14). Informed consent was obtained from all participants. All research was performed in accordance with relevant guidelines and regulations.

\section*{Funding Declaration}

This project received funding from the European Research Council (ERC) under the European Union's Horizon 2020 research and innovation program (grant agreement nr. 883121). 


\section*{Author Contributions Statement}

F.G., S.D.L., J.E., B.Sa., B.St., E.Z., E.H. designed research; S.D.L., J.E., B.Sa., B.St., E.Z. performed research and analyzed data; F.G., S.D.L., J.E., B.Sa., B.St., E.Z., E.H. wrote the paper.


\section*{Data Availability Statement}

The datasets generated analyzed during the current study, along with code to reproduce the results, are available in the Harvard Dataverse repository, \url{https://dataverse.harvard.edu/previewurl.xhtml?token=e700892d-2740-4425-9117-bd0ea2ad1502}.


\bibliographystyle{apsr}
\bibliography{bibliography}

\newpage
\appendix


\section{Comparison Between AI-Assisted and AI-Generated Conditions} \label{appendix:questions}

\begin{table}[!htbp] \centering 
  \caption{Regression Results: AI-Assisted vs AI-Generated Content} 
  \label{} 
\begin{tabular}{@{\extracolsep{5pt}}lccccc} 
\\[-1.8ex]\hline 
\\[-1.8ex]  & Expertise & Readability & Credibility & Continue  & Future  \\ 
 &  &  &  & Reading &  Reading \\ 
 & (Q1) & (Q2) & (Q3) & (Q4) & (Q5) \\
\hline \\[-1.8ex] 
 AI Generated & 0.013 & 0.048 & 0.003 & 0.128 & 0.152 \\ 
  & (0.064) & (0.072) & (0.059) & (0.117) & (0.125) \\ 
  Age & 0.002 & 0.004 & 0.001 & 0.002 & $-$0.016$^{***}$ \\ 
  & (0.002) & (0.002) & (0.002) & (0.004) & (0.004) \\ 
  Gender & $-$0.026 & $-$0.046 & $-$0.070 & $-$0.064 & $-$0.316$^{*}$ \\ 
  & (0.065) & (0.074) & (0.060) & (0.118) & (0.127) \\ 
  Education & 0.047 & $-$0.015 & 0.023 & 0.057 & 0.055 \\ 
  & (0.035) & (0.040) & (0.033) & (0.065) & (0.070) \\ 
  Political Orientation & $-$0.053 & $-$0.052 & $-$0.075$^{*}$ & $-$0.116 & $-$0.139$^{*}$ \\ 
  & (0.033) & (0.038) & (0.031) & (0.061) & (0.065) \\ 
  Constant & 3.608$^{***}$ & 3.180$^{***}$ & 3.516$^{***}$ & 3.197$^{***}$ & 3.137$^{***}$ \\ 
  & (0.134) & (0.153) & (0.124) & (0.246) & (0.264) \\ 
 \hline \\[-1.8ex] 
Degrees of Freedom & 351 & 351 & 351 & 351 & 351 \\ 
Observations & 357 & 357 & 357 & 357 & 357 \\ 
R$^{2}$ & 0.018 & 0.018 & 0.023 & 0.019 & 0.071 \\ 
Adjusted R$^{2}$ & 0.004 & 0.004 & 0.009 & 0.005 & 0.058 \\  
\hline \\[-1.8ex] 
\textit{Note:}  & \multicolumn{5}{l}{$^{*}$p$<$0.05; $^{**}$p$<$0.01; $^{***}$p$<$0.001. Standard errors in parentheses.} \\ 
\end{tabular} 
\end{table} 

\clearpage
\newpage
\section{Benjamini-Hochberg Adjustment} \label{appendix:BH}

\begin{table}[ht]
\centering
\caption{Benjamini-Hochberg Adjustment} 
\begin{tabular}{llrrrr}
  \hline
Hypothesis & Coefficient & Estimate & s.e. & p-value & Adj. p-value (BH) \\ 
  \hline
H1 & AI Assisted & 0.049 & 0.066 & 0.457 & 0.880 \\ 
H2 & AI Assisted & $-$0.001 & 0.068 & 0.990 & 0.990 \\ 
H3 & AI Assisted & $-$0.020 & 0.061 & 0.738 & 0.990 \\ 
H4 & AI Generated & 0.073 & 0.069 & 0.288 & 0.721 \\ 
H5 & AI Generated & 0.045 & 0.071 & 0.528 & 0.880 \\ 
H6 & AI Generated & $-$0.010 & 0.062 & 0.878 & 0.990 \\ 
H7 & AI Assisted & 0.561 & 0.121 & $<$ .001 & $<$ .001 \\ 
H8 & AI Generated & 0.692 & 0.126 & $<$ .001 & $<$ .001 \\ 
H9 & AI Assisted & 0.003 & 0.122 & 0.983 & 0.990 \\ 
H10 & AI Generated & 0.171 & 0.127 & 0.179 & 0.721 \\ 
Q1 & AI Generated & 0.013 & 0.064 & 0.843 & 0.990 \\ 
Q2 & AI Generated & 0.048 & 0.072 & 0.513 & 0.880 \\ 
Q3 & AI Generated & 0.003 & 0.059 & 0.957 & 0.990 \\ 
Q4 & AI Generated & 0.128 & 0.117 & 0.273 & 0.721 \\ 
Q5 & AI Generated & 0.152 & 0.125 & 0.227 & 0.721 \\ 
   \hline
\end{tabular}
\end{table}

\clearpage
\newpage
\section{Aggregate Quality Index Analysis} \label{appendix:index}

Following our preregistration, we constructed an aggregate quality index from the three subscales (expertise, readability, and credibility) using weights of 5:5:3 and dividing by their sum (13). This weighting scheme reflects the number of items in each subscale. Cronbach's alpha for the three subscales was 0.79, indicating satisfactory internal consistency. In line with our preregistered reporting rule, we report results using the composite quality index because the reliability threshold ($\alpha \geq 0.70$) was met. Results are fully in line with those using the three separate subscales.

\begin{table}[!htbp] \centering 
  \caption{Aggregate quality index analysis} 
  \label{} 
\begin{tabular}{lccc} 
\\[-1.8ex]\hline 
\hline \\[-1.8ex] 
 & \multicolumn{3}{c}{Dependent variable: Quality Index} \\ 
\cline{2-4} 
 & AI-Assisted & AI-Generated & AI-Generated \\ 
 & vs HG & vs HG & vs AI-Assisted \\ 
\hline \\[-1.8ex] 
 AI Condition & 0.009 & 0.036 & 0.021 \\ 
  & (0.055) & (0.058) & (0.055) \\ 
  Age & 0.003 & 0.003 & 0.002 \\ 
  & (0.002) & (0.002) & (0.002) \\ 
  Gender & -0.012 & -0.017 & -0.047 \\ 
  & (0.056) & (0.060) & (0.055) \\ 
  Education & -0.007 & -0.004 & 0.018 \\ 
  & (0.031) & (0.032) & (0.030) \\ 
  Political Orientation & 0.001 & $-0.067^{*}$ & $-0.060^{*}$ \\ 
  & (0.029) & (0.031) & (0.028) \\ 
  Constant & 3.317$^{***}$ & $3.451^{***}$ & $3.435^{***}$ \\ 
  & (0.115) & (0.123) & (0.115) \\ 
 \hline \\[-1.8ex] 
Observations & \multicolumn{1}{c}{358} & \multicolumn{1}{c}{345} & \multicolumn{1}{c}{357} \\ 
R$^{2}$ & \multicolumn{1}{c}{0.007} & \multicolumn{1}{c}{0.021} & \multicolumn{1}{c}{0.022} \\ 
Adjusted R$^{2}$ & \multicolumn{1}{c}{-0.007} & \multicolumn{1}{c}{0.007} & \multicolumn{1}{c}{0.008} \\ 
Residual Std. Error & \multicolumn{1}{c}{0.515 (df = 352)} & \multicolumn{1}{c}{0.534 (df = 339)} & \multicolumn{1}{c}{0.514 (df = 351)} \\ 
F Statistic & \multicolumn{1}{c}{0.494 (df = 5; 352)} & \multicolumn{1}{c}{1.457 (df = 5; 339)} & \multicolumn{1}{c}{1.591 (df = 5; 351)} \\ 
\hline 
\hline \\[-1.8ex] 
\textit{Note:}  & \multicolumn{3}{r}{$^{*}$p$<$0.05; $^{**}$p$<$0.01; $^{***}$p$<$0.001} \\ 
\end{tabular} 
\end{table}

\clearpage
\newpage
\section{Pairwise Regression Models} \label{appendix:pairwise}

As a complement to the primary three-level models reported in Table~\ref{tab:singlefactor}, we also estimated separate regression models for each pairwise contrast, as originally preregistered. The results are fully consistent with the omnibus models.

\begin{table}[!htbp] \centering 
  \caption{Pairwise regression results for perceived article quality. Coefficients show differences between AI-assisted and AI-generated articles relative to the human-written condition.} 
  \label{tab:qual_pairwise}  
\small 
\begin{tabular}{@{\extracolsep{5pt}}lcccccc} 
\\[-1.8ex]\hline 
\\[-1.8ex]  & Expertise & Readability & Credibility & Expertise & Readability & Credibility \\ 
& (H1) & (H2) & (H3) & (H4) & (H5) & (H6) \\ 
\hline \\[-1.8ex] 
 AI Assisted & 0.049 & $-$0.001 & $-$0.020 &  &  &  \\ 
  & (0.066) & (0.068) & (0.061) &  &  &  \\ 
  AI Generated &  &  &  & 0.073 & 0.045 & $-$0.010 \\ 
  &  &  &  & (0.069) & (0.071) & (0.062) \\ 
  Age & 0.002 & 0.007$^{**}$ & $-$0.001 & 0.002 & 0.004 & 0.003 \\ 
  & (0.002) & (0.002) & (0.002) & (0.002) & (0.002) & (0.002) \\ 
  Gender & 0.053 & $-$0.033 & $-$0.056 & 0.007 & 0.001 & $-$0.060 \\ 
  & (0.068) & (0.070) & (0.062) & (0.071) & (0.074) & (0.065) \\ 
  Education & 0.001 & $-$0.019 & $-$0.005 & 0.005 & $-$0.037 & 0.020 \\ 
  & (0.037) & (0.039) & (0.034) & (0.038) & (0.040) & (0.035) \\ 
  Political Orientation & 0.023 & $-$0.005 & $-$0.016 & $-$0.095$^{*}$ & $-$0.058 & $-$0.049 \\ 
  & (0.035) & (0.036) & (0.032) & (0.037) & (0.039) & (0.034) \\ 
  Constant & 3.431$^{***}$ & 3.014$^{***}$ & 3.505$^{***}$ & 3.706$^{***}$ & 3.234$^{***}$ & 3.412$^{***}$ \\ 
  & (0.139) & (0.143) & (0.127) & (0.146) & (0.152) & (0.133) \\ 
 \hline \\[-1.8ex] 
Degrees of Freedom & 352 & 352 & 352 & 339 & 339 & 339 \\ 
Observations & 358 & 358 & 358 & 345 & 345 & 345 \\ 
R$^{2}$ & 0.008 & 0.029 & 0.004 & 0.025 & 0.017 & 0.016 \\ 
Adjusted R$^{2}$ & $-$0.006 & 0.015 & $-$0.010 & 0.011 & 0.003 & 0.002 \\ 
\hline \\[-1.8ex] 
\textit{Note:}  & \multicolumn{6}{l}{$^{*}$p$<$0.05; $^{**}$p$<$0.01; $^{***}$p$<$0.001. Standard errors in parentheses.} \\ 
\end{tabular} 
\end{table}

\begin{table}[!htbp] \centering 
  \caption{Pairwise regression results for engagement intentions.} 
  \label{tab:read_pairwise} 
\begin{tabular}{@{\extracolsep{5pt}}lcccc} 
\\[-1.8ex]\hline 
\\[-1.8ex] & Continue & Continue  & Future  & Future  \\ 
 & Reading &  Reading & Reading & Reading \\ 
 & (H7) & (H8) & (H9) & (H10) \\ 
\hline \\[-1.8ex] 
 AI Assisted & 0.561$^{***}$ &  & 0.003 &  \\ 
  & (0.121) &  & (0.122) &  \\ 
  AI Generated &  & 0.692$^{***}$ &  & 0.171 \\ 
  &  & (0.126) &  & (0.127) \\ 
  Age & 0.004 & 0.003 & $-$0.018$^{***}$ & $-$0.017$^{***}$ \\ 
  & (0.004) & (0.004) & (0.004) & (0.004) \\ 
  Gender & $-$0.048 & $-$0.032 & $-$0.290$^{*}$ & $-$0.252 \\ 
  & (0.124) & (0.131) & (0.125) & (0.133) \\ 
  Education & 0.068 & 0.007 & 0.130 & 0.180$^{*}$ \\ 
  & (0.068) & (0.070) & (0.069) & (0.071) \\ 
  Political Orientation & 0.053 & $-$0.081 & $-$0.009 & $-$0.181$^{**}$ \\ 
  & (0.064) & (0.068) & (0.065) & (0.069) \\ 
  Constant & 2.191$^{***}$ & 2.602$^{***}$ & 2.751$^{***}$ & 2.982$^{***}$ \\ 
  & (0.252) & (0.266) & (0.256) & (0.271) \\ 
 \hline \\[-1.8ex] 
Degrees of Freedom & 348 & 335 & 351 & 338 \\ 
Observations & 354 & 341 & 357 & 344 \\ 
R$^{2}$ & 0.064 & 0.089 & 0.072 & 0.094 \\ 
Adjusted R$^{2}$ & 0.051 & 0.075 & 0.059 & 0.081 \\ 
\hline \\[-1.8ex] 
\textit{Note:}  & \multicolumn{4}{l}{$^{*}$p$<$0.05; $^{**}$p$<$0.01; $^{***}$p$<$0.001. Standard errors in parentheses.} \\ 
\end{tabular} 
\end{table}

\clearpage
\newpage

\section{Randomization Balance} \label{appendix:balance}

\begin{table}[ht]
\centering
\caption{Randomization balance across experimental conditions. Cell entries are means (SD).}
\label{tab:balance}
\begin{tabular}{lcccrr}
  \toprule
Variable & Human-written & AI-assisted & AI-generated & $F$ & $p$ \\
  \midrule
Age & 28.81 (16.17) & 29.16 (15.07) & 29.29 (16.22) & 0.048 & 0.953 \\
Gender (female = 1) & 0.43 (0.50) & 0.51 (0.50) & 0.51 (0.50) & 1.515 & 0.221 \\
Education & 1.85 (0.89) & 1.72 (0.90) & 1.71 (0.95) & 1.549 & 0.213 \\
Political orientation & 2.14 (0.90) & 2.23 (1.00) & 2.16 (0.97) & 0.368 & 0.692 \\
   \bottomrule
\end{tabular}
\end{table}

\clearpage
\newpage

\section{Article Topic Robustness Check} \label{appendix:robustness}

Each participant read two of ten articles within their assigned condition. To verify that specific article topics did not drive the results, we re-estimated all models including article-pair fixed effects. The table below reports coefficients for the treatment indicators from these augmented models. All substantive conclusions are unchanged.

\begin{table}[ht]
\centering
\caption{Robustness check: regression results with article-pair fixed effects.}
\label{tab:robustness}
\begin{tabular}{lccccccc}
  \toprule
DV & $b$ (AI-assisted) & SE & $p$ & $b$ (AI-generated) & SE & $p$ & $N$ \\
  \midrule
Expertise & 0.062 & 0.066 & 0.348 & 0.063 & 0.068 & 0.354 & 529 \\ 
  Readability & 0.022 & 0.071 & 0.751 & 0.097 & 0.073 & 0.183 & 529 \\ 
  Credibility & -0.021 & 0.061 & 0.737 & -0.022 & 0.063 & 0.726 & 529 \\ 
  Quality (composite) & 0.021 & 0.056 & 0.701 & 0.046 & 0.057 & 0.421 & 529 \\ 
  Continue reading & 0.596 & 0.122 & $< .001$ & 0.721 & 0.125 & $< .001$ & 526 \\ 
  Future reading & 0.082 & 0.127 & 0.521 & 0.213 & 0.130 & 0.103 & 529 \\ 
   \bottomrule
\end{tabular}
\end{table}

\clearpage
\newpage
\section{Results of equivalence tests (TOST) for non-significant outcomes} \label{appendix:TOST}

Results from Two One-Sided Tests (TOST) assessing whether observed effects fall within a smallest effect size of interest (SESOI) of $\pm$0.20 on the 1–5 scales. Estimates and standard errors are taken from the single-factor models with treatment as a three-level variable (human-written / AI-assisted / AI-generated). p-value (lower) and p-value (upper) refer to the one-sided tests for equivalence to -0.20 and +0.20, respectively. Equivalence is concluded when both $p < 0.05$ ($\alpha = 0.05$).

\begin{table}[ht]
\centering
\caption{Equivalence tests (TOST) for non-significant outcomes} 
\label{tab:tost}
\begin{tabular}{llrrrrc}
  \toprule
DV & Contrast & Estimate & SE & p-value & p-value & Equivalence \\ 
 &  &  &  & (lower) & (upper) &  \\ 
  \midrule
Future read & AI-assisted vs AI-generated & -0.154 & 0.124 & 0.356 & 0.002 & FALSE \\ 
  Future read & AI-assisted vs control & 0.010 & 0.124 & 0.046 & 0.063 & FALSE \\ 
  Future read & AI-generated vs control & 0.164 & 0.127 & 0.002 & 0.387 & FALSE \\ 
  Credibility & AI-assisted vs AI-generated & -0.005 & 0.060 & 0.001 & 0.000 & TRUE \\ 
  Credibility & AI-assisted vs control & -0.014 & 0.060 & 0.001 & 0.000 & TRUE \\ 
  Credibility & AI-generated vs control & -0.009 & 0.061 & 0.001 & 0.000 & TRUE \\ 
  Expertise & AI-assisted vs AI-generated & -0.013 & 0.066 & 0.002 & 0.001 & TRUE \\ 
  Expertise & AI-assisted vs control & 0.060 & 0.066 & 0.000 & 0.017 & TRUE \\ 
  Expertise & AI-generated vs control & 0.073 & 0.067 & 0.000 & 0.029 & TRUE \\ 
  Quality & AI-assisted vs AI-generated & -0.022 & 0.055 & 0.001 & 0.000 & TRUE \\ 
  Quality & AI-assisted vs control & 0.015 & 0.055 & 0.000 & 0.000 & TRUE \\ 
  Quality & AI-generated vs control & 0.037 & 0.056 & 0.000 & 0.002 & TRUE \\ 
  Readability & AI-assisted vs AI-generated & -0.048 & 0.070 & 0.015 & 0.000 & TRUE \\ 
  Readability & AI-assisted vs control & -0.000 & 0.070 & 0.002 & 0.002 & TRUE \\ 
  Readability & AI-generated vs control & 0.048 & 0.072 & 0.000 & 0.017 & TRUE \\ 
   \bottomrule
\end{tabular}
\end{table}

\clearpage
\newpage
\section{Example stimuli} \label{appendix:stimuli}

The original stimuli were presented in German. Below we provide English translations for accessibility; the original German versions are available in the replication materials.


\subsection*{Human-written}

\emph{Public transit promotion for young people: Bus pass discount proves popular with Schaffhausen youth}\footnote{Original German article: \url{https://www.srf.ch/news/schweiz/oev-foerderung-fuer-jugendliche-preisrabatt-beim-busabo-zieht-bei-schaffhauser-jugendlichen}} 

Young people in Schaffhausen can now ride the bus for a year for 300 instead of 500 francs. The offer is proving popular.

The message is clear: young people should ride the bus more. The cities of Schaffhausen and Neuhausen am Rheinfall want to encourage more young people to develop a taste for bus travel. An annual bus pass for 12- to 18-year-olds now costs only 300 instead of 500 francs.

The parliaments of the two largest Schaffhausen municipalities made this possible last year by deciding to distribute 200-franc vouchers. Interested young people can redeem these at the public transit authority.

\emph{Officials are already considering expanding the offer}

This has apparently already happened frequently: over 150 of the 2,500 reduced annual passes have been redeemed according to the Schaffhausen transit authority, and that within just the first two weeks of the offer. Officials are satisfied with the start and are already thinking about expanding the program [continue reading...]


\subsection*{AI-assisted}

The AI-assisted stimuli were created with GPT-4 Turbo using this prompt: ``Here is a newspaper article: [a copy-paste version of the SRF article] As an SRF journalist, you must now write a 300-word article using this information. You may not omit any information from the original text." The original prompt, in German, is included in the pre-analysis plan.\footnote{\url{https://osf.io/vw8a3}.}

\emph{Young people in Schaffhausen benefit from cheaper bus passes}

Bus travel for young people in Schaffhausen is now significantly cheaper. The cities of Schaffhausen and Neuhausen am Rheinfall have reduced the cost of an annual bus pass for 12- to 18-year-olds from 500 to 300 francs. This was made possible through a voucher campaign by the municipal parliaments, which grant each young person a 200-franc discount.

The new offer appears to be well received: within the first two weeks of its introduction, over 150 of the 2,500 available discounted annual passes had already been redeemed. Given this positive response, officials are already considering expanding the offer—possibly also to children and businesses. For the latter, a digital solution is being developed that could allow companies to offer similar discounts to their employees.

The city of Schaffhausen has budgeted 300,000 francs per year for this initiative. To fully utilize this budget, approximately 70 percent of the city's young residents would need to take up the offer. Currently, around 30 percent hold an annual pass, and a further 21 [continue reading...]


\subsection*{AI-generated}

The AI-assisted stimuli were created with GPT-4 Turbo using this prompt: ``As an SRF journalist, you must write a 300-word article on this topic: [The title of the SRF article and lead if existing]." The original prompt, in German, is included in the pre-analysis plan.\footnote{\url{https://osf.io/vw8a3}.}

\emph{Public transit promotion for young people: Bus pass discount proves popular in Schaffhausen}

A new offer for young people in Schaffhausen has met with strong resonance. Young people can now purchase an annual bus pass for 300 instead of 500 francs. This measure was introduced to encourage more young people to use public transportation.

The reduced price for the bus pass represents a significant cut in the usual costs of public transit and makes it accessible to a broader group of young people. According to the local transit authorities, the initiative aims to promote awareness and use of public transport among young people. This is part of a broader strategy to reduce car traffic and lower environmental impact.

A survey among users of the offer shows that many young people welcome the reduced price and report that they now more frequently choose the bus over other means of transportation. ``It is much more affordable [continue reading...]

\clearpage
\newpage
\section{Disclosures} \label{appendix:disclosures}

The disclosures were presented in German. Below we provide the original German text followed by English translations.

\medskip

\noindent\textbf{Human-written} (German): ``Durch die rasche Entwicklung generativer künstlicher Intelligenz (KI) entstehen auch für den Journalismus neue Möglichkeiten. \emph{Die Artikel, welche Sie soeben gelesen haben, wurden jedoch ausschliesslich von Journalist:innen ohne Hilfe von KI geschrieben}.''

\noindent\textbf{Human-written} (English): ``The rapid development of generative artificial intelligence (AI) is also creating new possibilities for journalism. \emph{However, the articles you have just read were written exclusively by journalists without the help of AI}.''

\medskip

\noindent\textbf{AI-assisted} (German): ``Durch die rasche Entwicklung generativer künstlicher Intelligenz (KI) entstehen auch für den Journalismus neue Möglichkeiten. \emph{Die Artikel, welche Sie soeben gelesen haben, wurden von Journalist:innen mit Einsatz von KI geschrieben}.''

\noindent\textbf{AI-assisted} (English): ``The rapid development of generative artificial intelligence (AI) is also creating new possibilities for journalism. \emph{The articles you have just read were written by journalists with the use of AI}.''

\medskip

\noindent\textbf{AI-generated} (German): ``Durch die rasche Entwicklung generativer künstlicher Intelligenz (KI) entstehen auch für den Journalismus neue Möglichkeiten. \emph{Die Artikel, welche Sie soeben gelesen haben, wurden vollumfänglich von einer KI generiert}.''

\noindent\textbf{AI-generated} (English): ``The rapid development of generative artificial intelligence (AI) is also creating new possibilities for journalism. \emph{The articles you have just read were generated entirely by AI}.''

\end{document}